\documentclass{physeauth}
\usepackage{graphicx}
\usepackage{amsmath}
\usepackage{amssymb}

\begin{document}

\begin{frontmatter}

\title{Dephasing of qubits by the Schr\"{o}dinger cat}

\author[us]{J. Dajka \thanksref{thank1}},
\author[us]{M.  Mierzejewski},
\author[us]{J.  {\L}uczka},
\author[ag]{P.  H\"anggi}

\address[us]{Institute of Physics, University of Silesia, 40--07 Katowice, Poland}
\address[ag]{Institute of Physics, University of Augsburg,
D-86135 Augsburg, Germany}

\thanks[thank1]{
Corresponding author.
E-mail: dajka@phys.us.edu.pl}

\begin{abstract}
We study the dephasing of a single qubit coupled to a bosonic bath. In particular,
we investigate  the case when the bath
is initially prepared in a pure state known as the Schr\"{o}dinger cat.
In clear contradistinction to the time-evolution of  an initial coherent state,
the time evolutions of the purity and the coherence factor now depend on
the  particular choice of the Schr\"{o}dinger cat state.
We also demonstrate that the evolution of the entanglement of a two--qubit system
depends on the initial conditions in a similar way.
\end{abstract}

\end{frontmatter}


\section{Introduction}
Controlling the dynamics of open quantum systems is of crucial importance
for the quantum information processing \cite{control}.
As there is no general method for analyzing the non--Markovian reduced dynamics,
the exactly solvable models may provide important and unbiased results.
One of the examples is the dephasing model \cite{lucz,ali,unruh,palma,petruc} that
describes an idealized case when the quantum system does not exchange
the energy with its environment.
This model has recently been studied in the context of entanglement dynamics
\cite{my0,pra0,pra1,miszcz} and the geometric phases \cite{myfaz}.
In particular, it has been shown in Ref.  \cite{pra0}
that the entanglement can effectively be controlled
by an external {\it finite}
bosonic quantum system prepared in so-called  {\it non--classical} states
 \cite{sir}.

In this paper we study the complementary case when the {\it infinite}
bosonic system  is initially prepared in the Schr\"{o}dinger cat state.
For a finite bosonic system such a state is defined as a superposition
of two coherent states with the same amplitudes but with phases  shifted by
$\pi$ \cite{sir}.
Here we generalize this notion   to  the case of infinite dimensional systems composed of bath and system dynamics.
We show that
the reduced dynamics  of the qubit depends on a specific
choice of the initial Schr{\"o}dinger cat state.  This is  in clear contrast to the  situation
when the initial state is  purely  coherent.
It holds true not only for  purity
and  coherence of a single qubit but
also for entanglement of a two--qubit system.

Due to the decoherence phenomenon,
the assumed initial state of an infinite bosonic bath
is   inaccessible in the present experiments. However, the development
of experimental techniques allows one to manipulate and
control systems devised from an increasing number of particles \cite{haroche}.
Therefore, the results presented in this paper may serve as a starting
point for understanding of qubits coupled to large bosonic systems prepared
in a desired quantum state.
Our choice of the initial state is motivated by the fact that
multiple Schr{\"o}dinger cat states can accurately
approximate any quantum state \cite{jan1,jan2}.

\section{Model}
We  consider a qubit $Q$, which interacts with the  environment
$R$.   The   Hamiltonian of the total system  reads    \cite{lucz,ali}
\begin{eqnarray}\label{ham}
H= H_Q \otimes  {\mathbb{I}}_R +{\mathbb{I}}_Q  \otimes H_R +H_I,
\end{eqnarray}
where  ${\mathbb{I}}_Q$   and ${ \mathbb{I}}_R$  are   identity operators  in corresponding Hilbert spaces of the  qubit $Q$ and the environment $R$, respectively.
  The qubit Hamiltonian $H_Q$  is in the form
\begin{eqnarray}\label{hamQ}
 H_Q = \varepsilon S^z  \equiv  \varepsilon \left(|1\rangle\langle 1| -  |-1\rangle\langle -1|\right)  ,
\end{eqnarray}
where   the  canonical basis of the qubit is   $\{|1\rangle, |-1\rangle\}$
and  $\pm \varepsilon$ are the energy levels of the qubit.
When $Q$ represents a particle of spin $S=1/2$,  the energy  $\varepsilon$ is proportional to the
magnitude of the external   magnetic field.
The  environment is assumed to be a boson field  described by the Hamiltonian
\begin{eqnarray}\label{hamR}
 H_R =
\int_0^\infty d\omega\;  h(\omega)  a^\dagger(\omega)a(\omega),
\end{eqnarray}
where the  real--valued dispersion relation   $h(\omega)$  specifies the environment,
e.g., $h(\omega) = \omega$  describes phonon or photon environment.
The operators
$ a^\dagger(\omega)$ and $a(\omega)$ are  the creation and annihilation boson operators, respectively.
The  coupling of the qubit to the environment is  described by the Hamiltonian
\begin{eqnarray} \label{int2}
H_{I}&=&|1\rangle\langle 1|\otimes H_+ +|-1\rangle\langle -1|\otimes H_{-},  \\ \mbox{with} \nonumber \\
H_\pm &=& \pm \int_0^\infty d\omega G(\omega) \left[ a(\omega) +a^\dagger(\omega)\right],
\end{eqnarray}
where the  function $G(\omega)$ is the coupling strength.
Without loosing generality, we assume that it is a real function.
The Hamiltonian  (\ref{ham}) can be rewritten in the form
  \begin{eqnarray} \label{H}
H=|1\rangle\langle 1|\otimes H_{1} +|-1\rangle\langle -1|\otimes H_{-1},   \\
H_{\pm1} =  H_R +   H_\pm  \pm \varepsilon. 
\label{Hpm}
\end{eqnarray}
Since there is no energy exchange (i.e. we use a non-demolition coupling) between the qubit and the environment,
our modeling corresponds to pure dephasing.
Hamiltonians   like (\ref{H}) have been exploited for description of
the  inter-conversion of electronic and vibrational energy  \cite{lin},    the electron--transfer
reactions  \cite{tang},   a quantum kicked rotator \cite{zoller97},  chaotic dynamics of a
periodically driven superconducting single electron transistor \cite{monta}
and  the Josephson flux qubit \cite{pozzo}, to mention but a few.

\section{Exact reduced dynamics }

The model we study is exactly solvable \cite{lucz,unruh,petruc},  i.e.,  the Schr\"odinger equation for the wave function $|\Psi(t)\rangle $ of the total system can be solved exactly.
Here we follow the method presented in  Ref. \cite{pra1}.
First, one needs to specify an  initial state $|\Psi(0)\rangle$.  Let us assume that   at the initial time $t=0$, the wave function has the form
\begin{eqnarray}\label{full_init}
|\Psi(0)\rangle=\left(b_1 |1\rangle + b_{-1} |-1\rangle\right)
\otimes|R\rangle,
\end{eqnarray}
where $b_1$ and $b_{-1}$ determine the qubit initial state and $|R\rangle$ is  the initial state
of the environment. Then
\begin{eqnarray}
\label{cohevol2}
|\Psi(t)\rangle= 
b_1 |1\rangle\otimes   |\psi_1(t)\rangle
+ b_{-1} |-1\rangle\otimes   |\psi_{-1}(t)\rangle,
\end{eqnarray}
where  $|\psi_i(t)\rangle = \exp[-H_i t]   |R\rangle$  ($i=\pm1$)
 can be rewritten in the form  \cite{pra1}
%
\begin{eqnarray}\label{cohevol}\
|\psi_1(t)\rangle &=&  \mbox{e}^{-i\Lambda_{1}(t)}
 D(g^{+}_t - g^{+} ) \; \mbox{e}^{-iH_R t } |R\rangle,
\nonumber\\
|\psi_{-1} (t)\rangle &=&  \mbox{e}^{-i\Lambda_{-1}(t)}
 D( g^{-} - g^{-}_t ) \; \mbox{e}^{-iH_R t } |R\rangle.
\end{eqnarray}
%
The  phases  $\Lambda_1(t)$ and $\Lambda_{-1}(t)$ are  given by
\begin{eqnarray}\label{L}
\Lambda_{1/-1}(t)= \pm \varepsilon t -\int_0^\infty d\omega g^2(\omega)
\left\{h(\omega) t-\sin [h(\omega)t] \right\},       \nonumber \\
\end{eqnarray}
where the abbreviation
$
g(\omega)= G(\omega)/h(\omega)
$
has been introduced.
For any  function $f$, the notation  $f_t$  stands for
\begin{eqnarray}
f_t(\omega)=e^{-ih(\omega) t}f(\omega).
\end{eqnarray}
For an arbitrary square--integrable function $f$,
the displacement operator $D(f) $ is defined as below \cite{brat}
\begin{eqnarray}\label{displacement}
D(f)=\exp\left\{\int_0^\infty d\omega \left[ f(\omega)a^{\dagger}(\omega) -  f^*(\omega)a(\omega)\right]\right\}.
\end{eqnarray}

The reduced qubit dynamics can be  obtained for  any factorisable  initial state of the form
\begin{eqnarray}
\varrho(0)=\sum_{i,j=1,-1} p_{ij}|i\rangle\langle j|\otimes| R\rangle
\langle R|,
\end{eqnarray}
where   $\varrho(0)$  is the initial statistical operator of the total system and   $p_{ij}$ are
non--negative parameters.
The reduced statistical operator  $\rho(t)$  for the qubit alone can be obtained by tracing the  environment degrees of freedom, namely,
\begin{eqnarray} \label{red}
 \rho(t) &=& \mbox{Tr}_R\left[ \varrho(t)\right]    \nonumber\\
 &=& \sum_{i,j=1,-1} p_{ij}|i\rangle \langle j|\otimes
\mbox{Tr}_R \left(e^{-iH_{i}t}|R\rangle \langle R |e^{iH_{j}t}\right)
\nonumber\\
&=& \sum_{i,j=1,-1} p_{ij} c_{ji}(t) |i\rangle \langle j|,
\label{doda}
 \end{eqnarray}
where $\mbox{Tr}_R$ denotes  the partial tracing over the environment variables, $H_{i}$ for
$i=\pm 1$ is given by Eq. (\ref{Hpm})   and
$c_{ji}(t) =  \langle\psi_j(t)|\psi_i(t)\rangle$
 is a scalar product between the functions
$|\psi_j(t)\rangle$ and $|\psi_i(t)\rangle$ in the environmental Hilbert space.
The initial state of the qubit  $|\theta, \phi\rangle$  is commonly parametrized by two angles on
 the Bloch sphere:  The
 polar angle $\theta$ and azimuthal angle $\phi$.  Then
\begin{eqnarray}\label{init}
|\theta, \phi\rangle =\cos(\theta/2)|1\rangle+\mbox{e}^{i\phi}\sin(\theta/2)|-1\rangle.
\end{eqnarray}
In this parametrization $b_1=\cos(\theta/2)$ and
$b_{-1}= \mbox{e}^{i\phi} \sin(\theta/2)$  (see  Eq. (\ref{full_init})) and
the initial density matrix  $\rho(0)$  of  the reduced qubit dynamics reads
\begin{eqnarray}\label{mat0}
\rho(0)=
\begin{pmatrix}
\cos^2(\theta/2)& (1/2)  \sin \theta\mbox{e}^{-i\phi} \\
 (1/2)  \sin \theta\mbox{e}^{i\phi} & \sin^2(\theta/2)
\end{pmatrix}.
 \end{eqnarray}
From Eq. (\ref{red})  we obtain  the density matrix  $\rho(t)$  in  the form
\begin{eqnarray}\label{mat1}
\rho(t)=
\begin{pmatrix}
\cos^2(\theta/2)& (1/2) A(t) \sin \theta\mbox{e}^{-i\phi} \\
 (1/2) A^*(t) \sin \theta\mbox{e}^{i\phi} & \sin^2(\theta/2)
\end{pmatrix}.
 \end{eqnarray}
All information about  influence of the environment   on the qubit  is  incorporated  in the dephasing function
$A(t) = c_{-1,1}(t)$. 

In the following we assume that initially the environment is in the pure
Schr\"{o}dinger cat state, which is defined by the relation
\begin{eqnarray}\label{kici}
|R\rangle=\frac{1}{\sqrt{N}}\left[|\alpha\rangle+\mbox{e}^{i\Phi}|-\alpha\rangle\right],
\end{eqnarray}
where 
$|\alpha \rangle=D(\alpha) | \Omega  \rangle $ is the  coherent state determined by the function $\alpha=\alpha(\omega)$ and   $|\Omega \rangle$ is the vacuum state of the bosonic bath. 
The normalization constant 
\begin{eqnarray}\label{Norm}
N=2+2 \cos(\Phi)\exp\left[-2 \int_0^{\infty} d {\omega} |\alpha(\omega)|^2 \right]. 
\end{eqnarray}
The phase  $\Phi$  allows to manipulate  the initial state of the environment. 
 In this case,  the  dephasing function becomes
\begin{eqnarray}  \label{A}
A(t) &=&N^{-1} \left[ \langle \alpha_{-1}(t)|\alpha_{1}(t)\rangle+\langle \alpha_{-1}(t)|-\alpha_{1}(t)\rangle \mbox{e}^{i\Phi} \right. \nonumber \\ &+& \left. \langle -\alpha_{-1}(t)|\alpha_{1}(t)\rangle\mbox{e}^{-i\Phi}+\langle -\alpha_{-1}(t)|-\alpha_{1}(t)\rangle \right] \nonumber \\
\end{eqnarray}
with $| \alpha_{\pm 1}(t) \rangle =\exp(-i H_{\pm 1} t) |\alpha \rangle $.
For the sake of brevity we calculate  the explicit form  of the dephasing function $A(t)$ for the case of given coherent
states $|\alpha \rangle $  determined  by real functions $\alpha(\omega)$ only.  As a first main result we find
\begin{eqnarray}  \label{Acat}
A(t) = N^{-1} A_0(t) \mbox{e}^{ -2 i\varepsilon t} &&  \left\{ 
 A_{+}(t)\mbox{e}^{-i\Phi}+
A_{-}(t)\mbox{e}^{i\Phi}
\right.  \nonumber\\
 && \left. + 2 \cos[4\Lambda_\alpha(t)] \right\},
\end{eqnarray}
where
\begin{eqnarray}  \label{A}
\Lambda_\alpha(t)&=&\int_0^\infty d\omega \alpha(\omega)g(\omega)\sin\left(h(\omega) t\right),  \\
\label{A0}
A_0(t) &=& \exp \left\{ -4\int_0^\infty d\omega g^2(\omega)[1-\cos(h(\omega) t)]	
 \right\},          \\
A_\pm(t) &=& \exp \left\{ - 2 \int_0^\infty d\omega \alpha^2(\omega) \right. \nonumber \\
&& \left. \mp 4 \int_0^\infty d\omega
  \alpha(\omega) g(\omega) [1-\cos(h(\omega) t)] \right\}.
 \label{Apm}
\end{eqnarray}
As we show next, the dephasing function $A(t)$ determines certain  quantifiers describing
various aspects of quantum information.

%
\section{Purity and coherence}

We start with basic quantifiers describing   the information loss of the qubit.  The first one  is
the   {\it purity} defined by:
\begin{eqnarray} \label{pur}
\mathcal{P}(t)=\mbox{Tr}(\rho^2(t))=\frac{1}{2}  \left(|A(t)|^2 -1 \right)  \sin^2\theta +1.
\end{eqnarray}
Its  interpretation is clear: The environment results in a decrease of  the purity. It is equal to $1$ for pure states and $1/2$ for maximally mixed states.
To quantify coherence, we introduce the coherence factor $\mathcal{C}(t)$ which is   determined by the evolution of the non--diagonal  elements of the qubit reduced density matrix,
\begin{eqnarray} \label{cohC}
|\rho_{12}(t)| = \mathcal{C}(t) |\rho_{12}(0)|.
\end{eqnarray}
Comparison of (\ref{mat0}) and (\ref{mat1}) yields
\begin{eqnarray} \label{coh}
 \mathcal{C}(t) =  |A(t)|.
\end{eqnarray}
The coherence factor is maximal in the absence of the qubit--bath interaction, i.e.,
$ \mathcal{C}(t)=1$,  and vanishes for the case of complete decoherence,
$ \mathcal{C}(t)=0$ .


\section{Entanglement decay}

In order to study the influence of the dephasing on the quantum  non--locality, we extend
the previous model and include  a second, completely independent, qubit $q$.
The Hamiltonian of such a composite system thus reads:
\begin{eqnarray}\label{hamer}
H&=&\left[H_{Q} + H_R +H_{I}\right]\otimes {\mathbb{I}}_{q}+H_{q},    \\
H_{q}&=& {\mathbb{I}}_{Q}\otimes {\mathbb{I}}_{R}\otimes \epsilon S_q^z.
\end{eqnarray}
We assume that the correlations between both the qubits are encoded in their initial entanglement.
For simplicity, we take the depolarized Bell states as the initial state, i.e.,
\begin{eqnarray}
 \rho(0)&=&(1-p)\rho_i+\frac{p}{4} {\mathbb{I}}_Q\otimes {\mathbb{I}}_q,\,\,\,\,\,\,\,\,\, i=1,\ldots,4
\end{eqnarray}
with
\begin{eqnarray}\label{AB}
 \rho_{1/2}&=&\frac{1}{2}\left[|-1, 1\rangle\pm|1, -1\rangle\right]\left[\langle -1, 1|\pm\langle 1, -1|\right],     \\
 \rho_{3/4}&=&\frac{1}{2}\left[|-1, -1\rangle\pm|1, 1\rangle\right]\left[\langle -1, -1|\pm\langle 1, 1|\right].
\end{eqnarray}
The depolarization accounts for an imperfect preparation of the initial state.
%
%
%

In a general case, the state of an open system is mixed. To quantify its entanglement, several useful
measures have been proposed \cite{negat}. One of the  operational measures is the
negativity, defined by $N (\rho) = \max( 0, -\sum_i \lambda_i)$, where $\lambda_i$
are the negative eigenvalues of the partially
transposed density matrix of  two qubits \cite{peres}.
For  the model under consideration the negativity can straightforwardly be evaluated for
an arbitrary evolution time $t$. One obtains
\begin{eqnarray} \label{neg}
N(\rho(t))=\max\left(0, \frac{1-p}{2}|A(t)|-\frac{p}{4}\right).
\end{eqnarray}
The negativity is positive for an entangled mixed state,
whereas it vanishes for unentangled states. Moreover, it presents an entanglement monotone
and can be used to quantify the degree of entanglement.

\section{Discussion}

The main quantifiers like the purity (\ref{pur}), the coherence factor  (\ref{coh})
 or the negativity  (\ref{neg})
depend directly on the dephasing function $A(t)$.  Therefore, we start with discussing its properties in further detail.
The dephasing function
depends on the qubit--environment coupling
via the functions  $g(\omega)=G(\omega)/h(\omega)$ and $\alpha(\omega)$. The latter one
defines the initial Schr{\"o}dinger cat state.
For convenience, we can assume that both   functions are real.
We also introduce the new function  $J(\omega)\equiv  \omega^2 g^2(\omega)$.
Then, the comparison of the function $A_0(t)$ (see Eq. (\ref{A0}))
with the standard expression for the decoherence function
(see e.g. Eq. (4.51) in Ref. \cite{petruc}), allows one to
identify $J(\omega)$ as the spectral density.
In the literature, there are several examples of  $J(\omega)$ in use.
A frequently used one is the generalized Drude form defined by    \cite{lucz}
\begin{eqnarray}\label{J}
 J(\omega) = \lambda\;\omega^{1+\mu}\exp(-\omega/\omega_c),
\end{eqnarray}
where  $\mu>-1$ and $\omega_c$ is the cut-off  frequency.
The case $\mu \in(-1,0)$ corresponds to a sub-Ohmic,  $\mu=0$ to the conventional Ohmic and
 $\mu \in(0,\infty)$ to a super-Ohmic environment.

One can observe that the long-time limit is given by
\begin{eqnarray}
\label{A_0}
A_0= \lim_{t\to \infty} A_0(t) = \exp \left\{ -4\int_0^\infty d\omega  J(\omega)/\omega^2\right\}.
\end{eqnarray}
The integral in this expression is infinite for a sub-Ohmic and an Ohmic environment.
Then  $A_0=0$ and the dephasing
function  diminishes to zero, $\lim_{t\to\infty} A(t)=0$. Consequently, the purity  (\ref{pur}),   coherence factor (\ref{coh}) and negativity (\ref{neg}) asymptotically take on the  following asymptotic long-time values
\begin{eqnarray} \label{as}
\mathcal{P}= 1 - \frac{1}{2}  \sin^2\theta, \quad \mathcal{C}=0,
\quad N=0.
\end{eqnarray}
One can see that for the sub-Ohmic and Ohmic environments
all the quantifiers are independent of any particular choice of $|\alpha\rangle$.
It means that in the long--time regime
the qubit properties do not depend any longer on the  initial Schr{\"o}dinger cat state.
The super-Ohmic case is more intriguing because $A_0 > 0$.
As it follows from the expression for the normalization constant $N$ (see  Eq. (\ref{Norm})), the  function $\alpha(\omega)$ is square--integrable.
Starting from the Cauchy--Schwarz  inequality one can find that also the integrals
in (\ref{A}) and (\ref{Apm}) exist, are finite and their values depend on the function $\alpha(\omega)$. In consequence, the dephasing function depends on both 
$\alpha$ and $\Phi$, i.e. on the initial state of the environment. 
Therefore  all characteristics (\ref{pur}),   (\ref{coh}) and  (\ref{neg}) do depend on the initial environment state, provided the environment is {\it super-Ohmic}.

It is instructive to compare Eq. (\ref{Acat}) with
the dephasing function obtained for an initial, purely  coherent state $|R\rangle =|\alpha \rangle$
of the environment. In this case the dephasing function $A(t)$ becomes
\begin{equation}
A(t) = \exp \left( -2 i\varepsilon t\right) \exp[-4i\Lambda_\alpha(t)]  A_0(t).
\end{equation}
In clear contrast to the initial Schr{\"o}dinger cat state,
$|A(t)|$ now depends only on $A_0(t)$. This fact implies   that  the purity  (\ref{pur}),   the coherence factor (\ref{coh}) and the negativity (\ref{neg}) are independent of
the initial state of the environment also for a super--Ohmic bath.

\section{Conclusions}

Dephasing characteristics of qubits coupled to a bosonic environment and prepared
in a Schr\"{o}dinger cat state has been investigated.
The properties of the reduced dynamics, as reflected in the purity
and coherence factor, have been shown to exhibit an explicit phase-dependence $\Phi$ as a  parameter  of the Schr\"{o}dinger cat state. Qualitatively the same behavior has been obtained
for the entanglement feature, being quantified by the negativity.
The main  conclusion is the following:
If the initial state of the environment is the coherent (or vacuum) state
then the informational quantifiers do not depend on the initial state.
However, if the initial state is a linear combination of two coherent
states as  the Schr\"odinger cat state then such quantifiers
as  the purity,   coherence factor  or the negativity
do depend on the initial state, -- via the function $\alpha$ and the phase  $\Phi$ --, of the environment, at least in the short-to-intermediate time regime.
For  the super-Ohmic environment this result holds true also in the long--time limit. Moreover, the $\Phi$-dependence allows one to  selectively control the
dephasing characteristics and the entanglement characteristics.

\section*{Acknowledgement}
The work  supported by
the Polish Ministry of Science and Higher Education
under the grant N 202 131 32/3786. We also gratefully acknowledge (P.H.) the financial support by the German Excellence
Initiative via the ``Nanosystems Initiative Munich (NIM)'' and the  DFG through the collaborative research centre  SFB 631

\begin{thebibliography}{99}

\bibitem{control}
R.~Alicki,   Controlled open quantum systems Irreversible
Quantum Dynamics, Lecture Notes in Physics vol {\bf 622}, Springer,  Berlin, 2003.

\bibitem{lucz} J. {\L}uczka,   Physica A  { 167} (1990) 919.

\bibitem{ali}  H. Spohn, Comm. Math. Phys.  { 123} (1987) 277; \\
 R. Alicki,  Open Sys. \& Information Dyn. { 11} (2004) 53.

\bibitem{unruh}  G. W. Unruh,  Phys. Rev.  A  { 51} (1995)  992.

\bibitem{palma} G. M. Palma, K. A. Suominen, A. K. Ekert, Proc. R. Soc. Lond. A { 452}
 (1996) 567.

\bibitem{petruc} H. -P. Breuer, F. Petruccione,  The theory of open quantum systems,
Oxford Univ. Press, Oxfrord, 2002.

\bibitem{my0} J. Dajka, M. Mierzejewski, J. \L uczka, J. Phys. A: Math. Theor. { 40}
(2007) F879.


\bibitem{pra0} J. Dajka, M. Mierzejwski, J. \L uczka, Phys. Rev. A { 77} (2008) 042316.

\bibitem{pra1} J. Dajka, J. \L uczka, Phys. Rev. A  {77} (2008) 062303.
	
\bibitem{miszcz} R. Doll, M.  Wubs, P.  H\"anggi, and S. Kohler,  Europhys. Lett. {76} (2006) 547;\\
R. Doll, M.  Wubs, P.  H\"anggi, and S. Kohler, Phys. Rev. B {76} (2007) 045317.

\bibitem{myfaz} J. Dajka, M. Mierzejewski, J. \L uczka, J. Phys. A: Math. Theor. { 41} (2008) F012001; \\ J. Dajka, J. Luczka,  J. Phys. A: Math. Theor.  { 41} (2008)  F442001.

\bibitem{sir} C. C. Gerry and P. L. Knight, Introductory Quantum Optics,
 Cambridge University Press,  Cambridge 2006.

\bibitem{haroche} S. Haroche, M. Brune, J. -M. Raimond, Eur. Phys. J. Special Topics
 { 159} (2008)  19.

\bibitem{jan1} P. Domokos, I. Ianszky, P. Adam, T. Larsen, Quantum Opt. {6} (1994) 187.

\bibitem{jan2} J. Janszky, A. Petak, C. Sibilia, C. Bertolotti, P. Adam,
 Quantum Semiclass. Opt. {7} (1995) 145.

\bibitem{lin} J. J. Markham, Rev. Mod. Phys.  { 31} 1956 956; \\
               Sheng Hsien Lin,  J. Chem. Phys. { 44} (1966) 3759.

\bibitem{tang} J. Tang, Chem. Phys. { 188} (1994) 143.


\bibitem{zoller97} S. A. Gardiner, J. I. Cirac, P. Zoller, Phys. Rev. Lett. { 97} (1997) 4790.


\bibitem{monta}  S. Montangero, A. Romito, G. Benenti, R. Fazio, Europhys. Lett. { 71} (2005) 893.

\bibitem{pozzo}  E. N. Pozzo, D. Dominguez,  Phys. Rev. Lett. { 98} (2007) 057006.


\bibitem{brat} O. Brattelli,  D. W.  Robinson,
Operator Algebras and quantum statistical mechanics, Springer, Berlin, 1997.

\bibitem{negat} F. Verstraete, K. Audenaert, J. Dehaene, B. De Moor,
 J. Phys. A: Math. Gen. { 34} (2001) 10327;\\
 G. Vidal, R. F. Werner, Phys. Rev. A {65} (2002) 032314.

\bibitem{peres} A. Peres, Phys. Rev. Lett. { 77} (1996) 1413.

\end{thebibliography}

\end{document}